\documentclass{aa}

\usepackage{lineno}
\usepackage{amsmath}
\usepackage{hyperref}
\usepackage[varg]{txfonts}
\usepackage[utf8]{inputenc}
\usepackage{xspace}
\usepackage{graphicx}
\usepackage{color}

\hypersetup{%
  colorlinks=true,
  citecolor=blue,
  linkcolor=blue
} 
\usepackage{natbib}
\bibpunct{(}{)}{;}{a}{}{,} % to follow the A&A style

\newcommand{\muG}{$\mu$G\xspace}
\newcommand{\kms}{km\,s$^{-1}$\xspace}
\newcommand{\Msun}{$M_\odot$\xspace}
\newcommand{\god}{\textsc{Godunov-MHD}\xspace}

\begin{document}

\title{3D global simulations of a cosmic-ray-driven dynamo in dwarf galaxies}

\author{%
        H.~Siejkowski\inst{\ref{cyf},\ref{oa}} \and
        K.~Otmianowska-Mazur\inst{\ref{oa}} \and
        M.~Soida\inst{\ref{oa}} \and
        D.J.~Bomans\inst{\ref{rub},\ref{rubplasma}} \and
        M.~Hanasz\inst{\ref{umk}}
}

\institute{%
  AGH University of Science and Technology, ACC Cyfronet AGH, ul. Nawojki 11, 30-950 Krak\'ow 23,
  P.O.Box 386, Poland,\\
  \email{h.siejkowski@cyfronet.pl}\label{cyf}
  \and
  Astronomical Observatory of the Jagiellonian University, ul. Orla 171, 30-244 Krak\'ow,
  Poland\label{oa}
  \and
  Astronomical Institute of Ruhr-University Bochum, Univerist\"{a}tsstr. 150/NA7, D-44780 Bochum,
  Germany\label{rub}
  \and
  Ruhr-University Bochum Research Department: Plasmas with Complex Interactions,
  Univerist\"{a}tsstr. 150, D-44780 Bochum, Germany\label{rubplasma}
  \and
  Centre for Astronomy, Nicolaus Copernicus University, Faculty of Physics, Astronomy and
  Informatics, Grudziadzka 5, PL-87100 Toru{\'n}, Poland\label{umk}
}

\abstract% 
% context
{Star-forming dwarf galaxies can be seen as the local proxies of the high-redshift building blocks of more
massive galaxies according to the current paradigm of the hierarchical galaxy formation.  They are
low-mass objects, and therefore their rotation speed is very low. Several galaxies are observed to
show quite strong magnetic fields. These cases of strong ordered magnetic fields seem to correlate
with a high, but not extremely high, star formation rate.}
% aims
{We investigate whether these magnetic fields could be generated by the cosmic-ray-driven dynamo.
The environment of a dwarf galaxy is unfavourable for the large-scale dynamo action because of the very
slow rotation that is required to create the regular component of the magnetic field.}
% methods
{We built a 3D global model of a dwarf galaxy that consists of two gravitational components: the
stars and the dark-matter halo described by the purely phenomenological profile proposed previously.
We solved a system of magnetohydrodynamic (MHD) equations that include an additional cosmic-ray
component described by the fluid approximation.}
% results
{We found that the cosmic-ray-driven dynamo can amplify the magnetic field with an exponential
growth rate. The $e$-folding time is correlated with the initial rotation speed. The final mean
value of the azimuthal flux for our models is of the order of few \muG and the system reaches its
equipartition level. The results indicate that the cosmic-ray-driven dynamo is a process that can
explain the magnetic fields in dwarf galaxies.}
% conclusions
{}

\keywords{magnetohydrodynamics (MHD), dynamo, galaxies: magnetic fields, galaxies: dwarf, methods:
numerical}

\maketitle

\section{Introduction} 
\label{sec:introduction}

Star-forming dwarf galaxies are smaller, fainter, and less massive than their spiral counterparts,
but they are the most numerous population in the Universe \citep{blanton02}.  The magnetic fields in
dwarf galaxies may play a very important role. First, the observations show that the magnetic field
is an important source of pressure for the interstellar medium \citep{boulares90}. It is often
assumed that the interstellar medium (ISM) in galaxies is in equipartition with the almost equally
distributed energy in magnetic fields, cosmic rays, and turbulence. Second, these objects are less
massive, hence the gravitational potential well is shallower and this facilitates the escape of the
gas from the galaxy in the form of galactic winds \citep{maclow99}. This wind can also drag the
magnetic field out of the disk into the intergalactic medium (IGM) since the magnetic field is
frozen into the outflowing plasma. Possible magnetisation of the IGM via the magnetised wind has
been studied by \cite{bertone06} and for dwarf galaxies by \cite{kronberg99} and
\cite{scannapieco10}.  Local simulations of the cosmic-ray dynamo process in dwarf galaxies
\citep{siejkowski10} show significant loss of the magnetic field from the domain and it depends on
the supernova rate (SNR).  Studies of dwarf galaxy formation by \cite{dubois10} also implied IGM
seeding via galactic winds. Additionally, there are studies on galactic winds driven by cosmic rays
by \cite{booth13} and \cite{hanasz13}.

Strong magnetic fields were discovered in a bright dwarf irregular galaxy \object{NGC 4449} with a
total field strength of about 12~\muG and a regular component of up to 8~\muG
\citep{klein96,chyzy00}.  \cite{kepley10} reported similar magnetic fields in \object{NGC 1569}. The
radio observations of these two galaxies show some large-scale magnetic fields with a sign of a
spiral pattern, but no optical counterparts.  The rotation measure maps imply that the magnetic
field is almost parallel to the disk plane. In other dwarf objects such as \object{NGC 6822},
\object{IC 10} \citep{chyzy03} and the \object{Large Magellanic Cloud} \citep{klein89,gaensler05}
the observed magnetic fields are weaker, reaching a value about 5-7~\muG.  It is worth noting that
these galaxies, especially \object{NGC 1569} and \object{NGC 4449}, are under strong influence of
infalling gas from the surroundings, which have a significant impact on the magnetic-field
structure.

Observations of the above mentioned objects brought some insight into the dynamo process in dwarf
galaxies, but all these objects are optically bright and showed disturbed kinematics. Therefore the
sample might be influenced by strong selection effects. \cite{chyzy11} completed a sample of dwarf
and small irregular galaxies from the Local Group.  They found that the star formation rate
(SFR) regulates the strength of magnetic fields, because the supernova rate, which is proportional
to the SFR, contributes to drive the turbulence in the ISM\@. These results are very similar to the
conclusions of our previous theoretical study \citep{siejkowski10} of the cosmic-ray-driven dynamo
in the medium of irregular galaxies.  \cite{chyzy11} have also investigated a possible relation
between the maximum $v_\textrm{rot}$ and the magnetic-field strength. For slow rotation
$(<40~\textrm{km/s})$ all galaxies have weak fields, below 4~\muG.  With increasing 
maximum rotation speed in the following objects the observed magnetic fields are stronger.  However,
some objects, such as \object{NGC 4449}, show very strong magnetic fields, but their rotation is
rather slow.  The relation of the magnetic field versus the maximum rotation speed is probably
distorted by the fact that objects with strong magnetic fields are undergoing heavy star
formation. The contribution to the turbulent energy by supernovae explosions can cause strong
disturbances in the velocity pattern, therefore estimating the maximum rotation speed and
assigning a clear rotation curve is difficult. 

It is believed that the magnetic dynamo is responsible for the strength and structure of magnetic
fields in galaxies \citep{beck09}.  One of the recent dynamo models is driven by thermal energy
output from supernovae explosions, described by \cite{gressel08}. These authors found that the $e$-folding
time of the amplification mechanism is about $\tau_e = 250$~Myr and is dependent on the rotation
speed. Another model suggested by \cite{parker92} relies on the buoyancy instability of interstellar
medium filled with magnetic fields and cosmic rays \citep{parker65}.  The cosmic-ray-driven dynamo
was constructed for the first time within the framework of a local shearing box model by
\cite{hanasz04}.  An extensive parameter study of this local model was presented in
\cite{hanasz09param}.  The $e$-folding time-scale of the magnetic-field amplification by the
cosmic-ray-driven dynamo  is generally comparable to the galactic rotation period and can be as short
as 140~Myr \citep{hanasz06}.  \cite{siejkowski10} investigated the cosmic-ray driven dynamo action
in low-mass objects, such as dwarf- and intermediate-mass irregular galaxies.  This study led to
the following results: the growth rate of the magnetic field is strongly dependent on the rotation
speed, but for objects with $v_\textrm{rot} > 40$~km~s$^{-1}$ the saturation of the dynamo is
reached after the same period of time.  The $e$-folding time is also dependent on the SNR and the
time of quiescent state (no supernovae activity).  The larger the SNR, the faster the growth rate,
but excessive supernova activity can suppress the dynamo action.  \cite{hanasz09global} demonstrated
cosmic-ray-driven dynamo action via global disk spiral galaxies, and \cite{kulpa11} confirmed the
result for barred galaxies.  They found that the azimuthal magnetic flux grows on a time-scale of
about 270~Myr.

% section introduction (end)

\section{Numerical model and setup} 
\label{sec:numerical_model}

To simulate the dwarf galaxy model a numerical code called \god was employed \citep{kowal09}. It
solves the system of ideal magnetohydrodynamic (MHD) equations in a conservative form in 3D space.  The key
elements and assumptions of the cosmic-ray-driven dynamo global galactic model were adopted from
\cite{hanasz09global}. Below we describe our choice of simulation parameters.  An isothermal
equation of state was assumed, that is $p \equiv \rho c_s^2$, where $c_s$ is the isothermal speed of
sound  set to 7~km~s$^{-1}$, which corresponds to a gas temperature of 6\,000~K.  We
assumed the magnetic diffusivity $\eta$ to be constant and equal to 0.1~kpc$^{2}$~Gyr$^{-1} = 3 \times
10^{25}$~cm$^2$~s$^{-1}$. The investigation by \cite{hanasz09param} showed that this
value is optimal for the growth of the magnetic field in the buoyancy-driven dynamo.

The cosmic ray (CR) component is described by the diffusion-advection transport equation in terms of
fluid approximation following \cite{schlickeiser_lerche} and \cite{hanasz03cr}.  We related the CR
pressure to the CR energy density $e_\mathrm{cr}$ via the adiabatic CR index, that is $p_\mathrm{cr}
\equiv (\gamma_\mathrm{cr} - 1) e_\mathrm{cr}$ and $\gamma_\mathrm{cr} = 14/9$ adopted from
\cite{ryu03}.  Following \cite{giacalone99}, we assumed that the diffusion of cosmic rays is
anisotropic with respect to the direction of the magnetic field.  The typical values of the
diffusion coefficient found by modelling CR data \citep[see e.g.][]{strong07} are $(3\div5) \times
10^{28}$~cm$^{2}$~s$^{-1}$.  The applied value of the perpendicular CR diffusion coefficient is
$K_\perp = 1$~kpc$^2$~Gyr$^{-1} = 3 \times 10^{26}$~cm$^2$~s$^{-1}$ and the parallel one is
$K_\parallel = 10$~kpc$^2$~Myr$^{-1} = 3 \times 10^{27}$~cm$^2$~s$^{-1}$. The parallel diffusion
coefficient is 10\% of the realistic value because the time-step of the explicit diffusion algorithm
becomes prohibitively short when the diffusion coefficient is too high. The effect of the reduced CR
diffusion coefficients was investigated by \cite{hanasz09param}, showing that the magnetic-field
growth only slightly depends on the $K_\parallel$ value, but the key factor in the cosmic-ray-driven
dynamo is the anisotropy of the diffusion.

A single supernova explosion was modelled by a 3D Gaussian distribution of cosmic-ray energy input
and equals 10\% of the canonical kinetic energy output of the supernova explosion, that is
$10^{51}$~erg.  In the initial period in $t \in (100~\textrm{Myr},400~\textrm{Myr})$, every one in
ten explosions injects a randomly oriented dipole magnetic field into the ISM in addition to the CR
energy input.  The injection of the magnetic field is only to seed the dynamo action through a
random field at the beginning of the simulation. We stopped the injection because (1) it allowed us
to study the efficiency of the pure cosmic-ray-driven dynamo while the injected magnetic field was
only the seed field and (2) after several hundred Myr the seeding becomes insignificant with respect
to the existing field.

The position of each supernova explosion was chosen randomly with respect to the local gas density,
that is $\rho^{3/2}$. This follows the simple self-gravitational picture drawn by \citet{schmidt59}
and \citet{kennicutt89}, where the large-scale star formation rate volume density scales with
$\rho^{3/2}$. The number of supernovae is given by the supernova explosion frequency over the disk
area. In our model its value was set to $f_\mathrm{SN} = 3 \times 10^3$~kpc$^{-2}$~Gyr$^{-1}$ and it
was constant during whole simulation time. This value corresponds roughly to the density of the star
formation rate ($\Sigma$SFR) equal to $10^{-3}$~\Msun~yr$^{-1}$~kpc$^{-2}$ and is a typically found
in non-starbursting dwarf galaxies \citep[see e.g.][]{chyzy11}. 

\begin{table}
  \centering
  \caption{Parameters of the models v40 and v70.}
  \begin{tabular}{rccl}
    \hline \hline

    Parameter & v40  & v70  & Unit \\ \hline

    Mass of stars & 1.0  & 6.0  & $10^9$~\Msun \\
%    $a$       & 1.4  & 2.2  & kpc \\
 $\rho_0$     & 21.5 & 15.0 & $10^6$~\Msun~kpc$^{-3}$ \\
    $r_0$     & 1.4  & 2.4  & kpc \\
 $\rho_g$     & 29.5 & 29.5 & $10^6$~\Msun~kpc$^{-3}$ \\
    $R_c$     & 1.2  & 3.6  & kpc \\
    $v_\varphi^\textrm{max}$ & 40 & 70 & km~s$^{-1}$ \\

    \hline

  \end{tabular}
  %\tablefoot{The symbols meaning can be found in the Sec.~\ref{sec:numerical_model}.}
  \label{tab:params}

\end{table}

The dwarf galaxy potential well is given by two components: a dark-matter (DM) halo and a thin
stellar disk. This type of galaxy has no bulge \citep{governato10}, which is present in more massive
disk galaxies.  In our model the stars are distributed in an infinitesimally thin Kuzmin disk
following previous numerical works on dwarf galaxies \citep{marcolini03,marcolini04}.  For the DM
halo we used the purely phenomenological profile proposed by \cite{burkert95}. Its gravitational
potential is described by

\begin{align}
  \begin{split}
    \phi_\mathrm{DM}(r) = & -\pi G \rho_0 r_0^2 \left\{\pi
    - 2 (1 + x^{-1})\arctan{x}
    \right. \\ & \left. + 2 (1 + x^{-1}) \ln{(1+x)} \right. 
     \left. - (1 - x^{-1}) \ln{(1 + x^2)}
    \right\},
    \label{eq:burkert}
  \end{split}
\end{align}

\noindent where $G$ is the gravitational constant, $\rho_{0}$ is the central density, $r_0$ is the
core radius, $r$ is the distance to the centre, and $x \equiv r/r_0$.

The gas distribution of a dwarf galaxy was set  in hydrostatic equilibrium in its initial state. The
gas density distribution in equatorial plane was assumed in the following form:

\begin{equation}
  \rho(R,z=0)=\frac{\rho_g}{\left[1 + ({R}/{R_c})^2\right]^2},
  \label{eq:gas_dens}
\end{equation}

\noindent where $\rho_g$ and $R_C$ are the central gas density and core radius, respectively. To
find the global gas distribution we used the potential method described in \cite{wang10}.  The
detailed parameters of the gravitational potentials are shown in Tab.~\ref{tab:params} and its
values were mostly taken from \cite{marcolini03} with only slight modifications.  From the gas
distribution the CR component distribution was found, assuming that they are in pressure
equilibrium. We assumed $B=0$ at $t = 0$.  

The dwarf galaxy model was simulated in 3D Cartesian domain of size $14\times14\times7$~kpc in $x$,
$y$, $z$ coordinates, respectively. The cell size was $50$~pc in each direction. Upper and lower
boundary conditions were set to outflow, and we enforced the gas to follow the prescribed rotation
curve in the horizontal domain boundaries.

\section{Results and discussion} 
\label{sec:results}

\begin{figure}
  \includegraphics[width=0.48\textwidth]{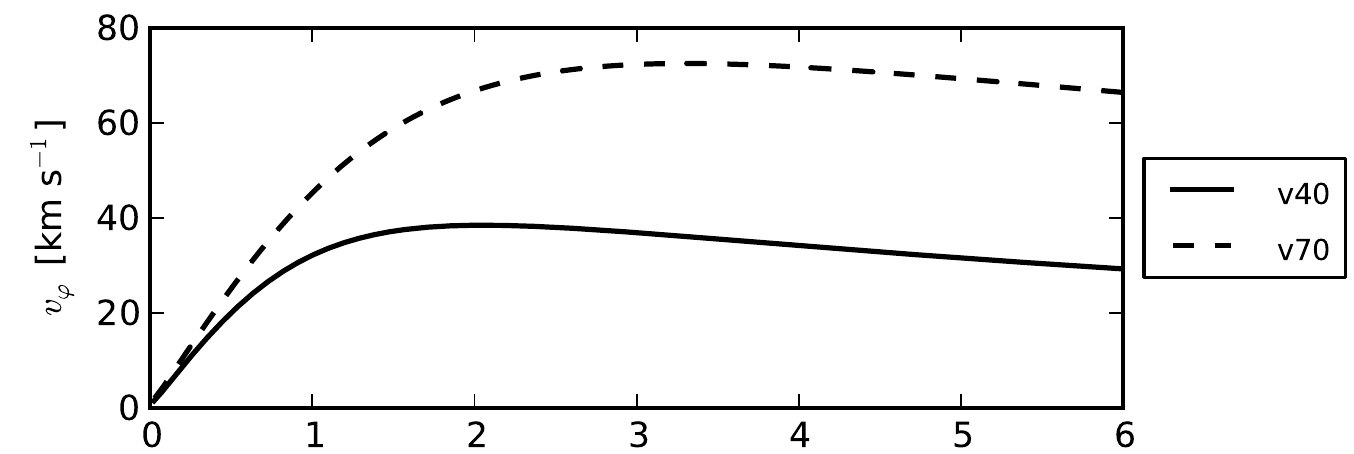}
  \caption{Rotation curves for models from Tab.~\ref{tab:params}.}
  \label{fig:rc}
\end{figure}

The models presented in Tab.~\ref{tab:params} correspond to the different maximum rotation speeds.
We set up models with $v_\varphi^\textrm{max} = 40$~km~s$^{-1}$ and 70~km~s$^{-1}$. The model v40 
corresponds to objects like \object{NGC 1569} or \object{IC 10}, while the v70 model mimics the
\object{Large Magellanic Cloud} or \object{NGC 6822} \citep[e.g.][]{chyzy11}. The rotation curves of these
models are shown in Figure~\ref{fig:rc}.

\begin{figure*}
  \includegraphics[width=0.99\textwidth]{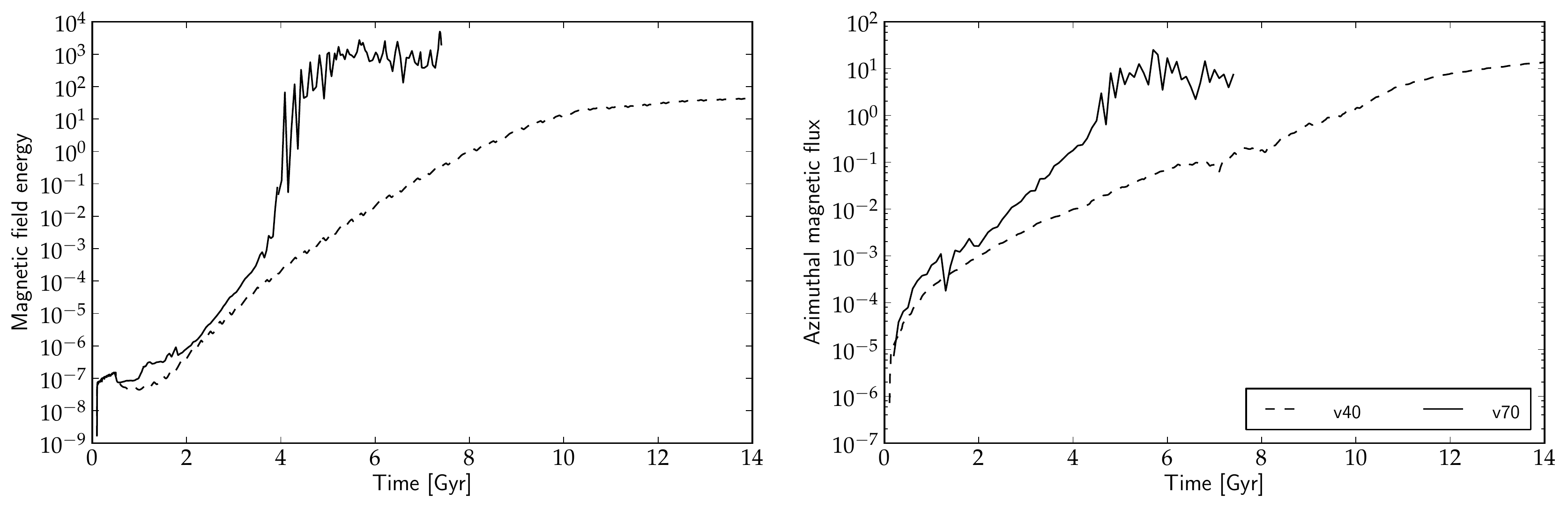}
  \caption{\textit{Left panel:} magnetic-field energy evolution for models from
  Tab.~\ref{tab:params}. \textit{Right panel:} total azimuthal magnetic flux evolution. The plotted
  value is the absolute value of $B_\phi$, because it changes its sign at the beginning of the
  simulation and therefore it is difficult to show it in a log-plot.}
  \label{fig:bevo}
\end{figure*}

The evolution of the total magnetic energy and the magnetic flux  are shown in
Figure~\ref{fig:bevo}. In both models the magnetic-field energy and the magnetic flux are growing
exponentially in time.   The initial peak, visible up to $t=400$~Myr, is caused by the magnetic field injected
into the system via the magnetised supernovae. Later the magnetic field injection was stopped, and
the magnetic growth was only caused by the operation of the dynamo. 

After the initial injection of the magnetic field the system experiences a period of
stabilization and the magnetic-field energy decreases by about half an order of magnitude. After
this, at about $t=1$~Gyr, the magnetic field starts to be amplified. In the evolution of the
magnetic flux we see a constant growth, without any stabilization periods. This is because the
injected magnetic field is random and is ordered by the global rotation. This also suggests that
during the stabilization period the randomly oriented magnetic field is expelled and/or dissipated
by magnetic diffusivity beyond the simulation domain, while the ordered azimuthal magnetic flux is
retained.

When the dynamo begins to operate effectively, the magnetic field and its azimuthal part are 
exponentially amplified in time. The growth rate of the azimuthal magnetic flux, measured by the
$e$-folding time, is different for each of the models and equals 1\,023~Myr for v40 and 458~Myr
for v70.  For model v70 the equipartition level is reached at about $t=5$~Gyr. Around
that time some peaks appear in the evolution of the magnetic-field energy, but when the equipartition
level is reached - they become damped. In model v40 the equipartition level is reached at
$t=10$~Gyr, but the transition is smoother than in the previous case. In both cases the saturation
of the growth of $B_\phi$ occurs at the same time as for the magnetic energy. 

\begin{table}
  \centering
  \caption{Comparison of the simulation results with observations.}
  \label{tab:obssim}

  \begin{tabular}{r|ccc}

      \hline \hline
      Model & $v_\phi^{\textrm{max}}$ & $\Sigma$SFR                  & $B$  \\
            & [km s$^{-1}$]           & [\Msun~yr$^{-1}$~kpc$^{-2}$] & [$\mu$G] \\
      \hline
      v40 & 40 & $10^{-3}$ & 1.0 \\
      v70 & 70 & $10^{-3}$ & 8.0 \\
      \hline
      Object & & & \\
      \hline
      IC 1613  & 37 & $3.7 \times 10^{-4}$ & $2.8 \pm 0.7$ \\
      NGC 4449 & 40 & $1.7 \times 10^{-2}$ & $9.3 \pm 2.0$ \\
      NGC 1569 & 42 & $1.5 \times 10^{-1}$ & $14.0 \pm 3.0$ \\
      IC 10    & 47 & $5.2 \times 10^{-2}$ & $9.7 \pm 2.0$ \\
      NGC 6822 & 60 & $6.0 \times 10^{-3}$ & $4.0 \pm 1.0$ \\
      LMC      & 72 & $4.0 \times 10^{-3}$ & $4.3 \pm 1.0$ \\
      \hline
        
  \end{tabular}

  \tablefoot{The last column shows the value of mean magnetic field in the disk ($\bar{B}$) for the
  models; for real objects the total magnetic field ($B_\textrm{tot}$) is given. All the
  observational values are taken from \cite{chyzy11}.} 
    
\end{table}

The final mean value of the magnetic field for model v40 is $\bar{B}(t=10~\textrm{Gyr}) = 1.0$~\muG
and for model v70 it is $\bar{B}(t=5~\textrm{Gyr}) = 8.0$~\muG.  The results are compared with the
magnetic field found in observations of dwarf galaxies in Table~\ref{tab:obssim}. The model v70 can
be compared with \object{NGC 6822} and the \object{Large Magellanic Cloud}, which rotate as fast as
60-70~\kms, and whose $\Sigma$SFR is of the order of $10^{-3}$~\Msun~yr$^{-1}$~kpc$^{-2}$. In both
cases the observed magnetic field is about 4~\muG. The model v40 with respect to its rotational
velocity can be compared with galaxies such as \object{IC 1613}, \object{IC 10}, \object{NGC 4449},
and \object{NGC 1569}. The observed magnetic field in the first object is about 2.8~\muG, which
agrees well with our simulations. In other objects the magnetic field is much higher, starting from
9~\muG up to even 14~\muG. However, these objects have a much higher SFR. In our modelling the
applied supernova frequency is equivalent to $\Sigma\textrm{SFR} =
10^{-3}$~\Msun~yr$^{-1}$~kpc$^{-2}$, while in these galaxies the observed values are given in
Table~\ref{tab:obssim}.  The $\Sigma$SFR value is comparable only for \object{IC 1613}, but in other
cases the observed values are much higher.  This leads to the conclusion that the enhanced SFR can
increase the magnetic field in the galaxy disk, but to state it more clearly, more numerical
investigations need to be made.

\begin{figure*}
  \centering
  \includegraphics[width=0.32\textwidth]{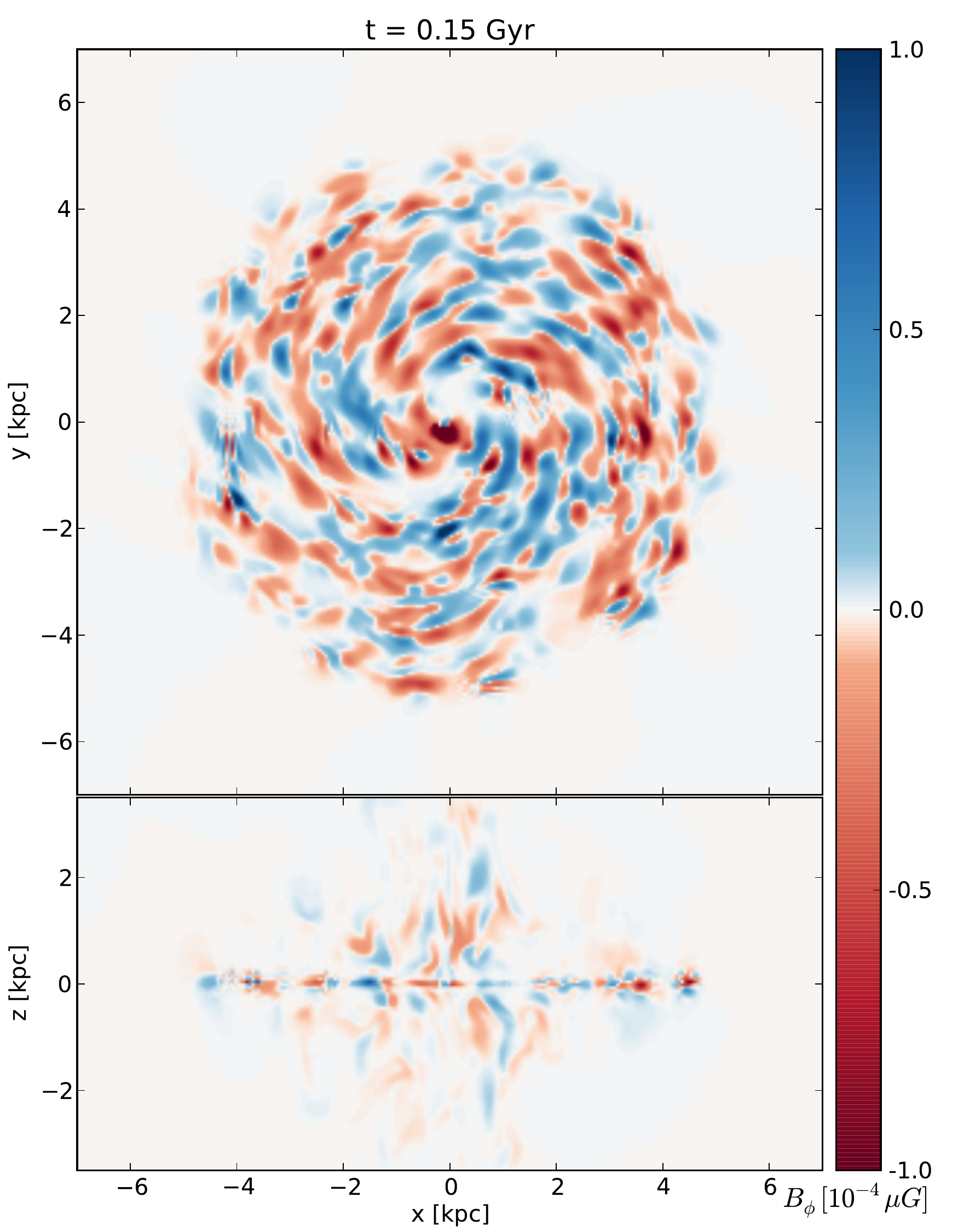}
  \includegraphics[width=0.32\textwidth]{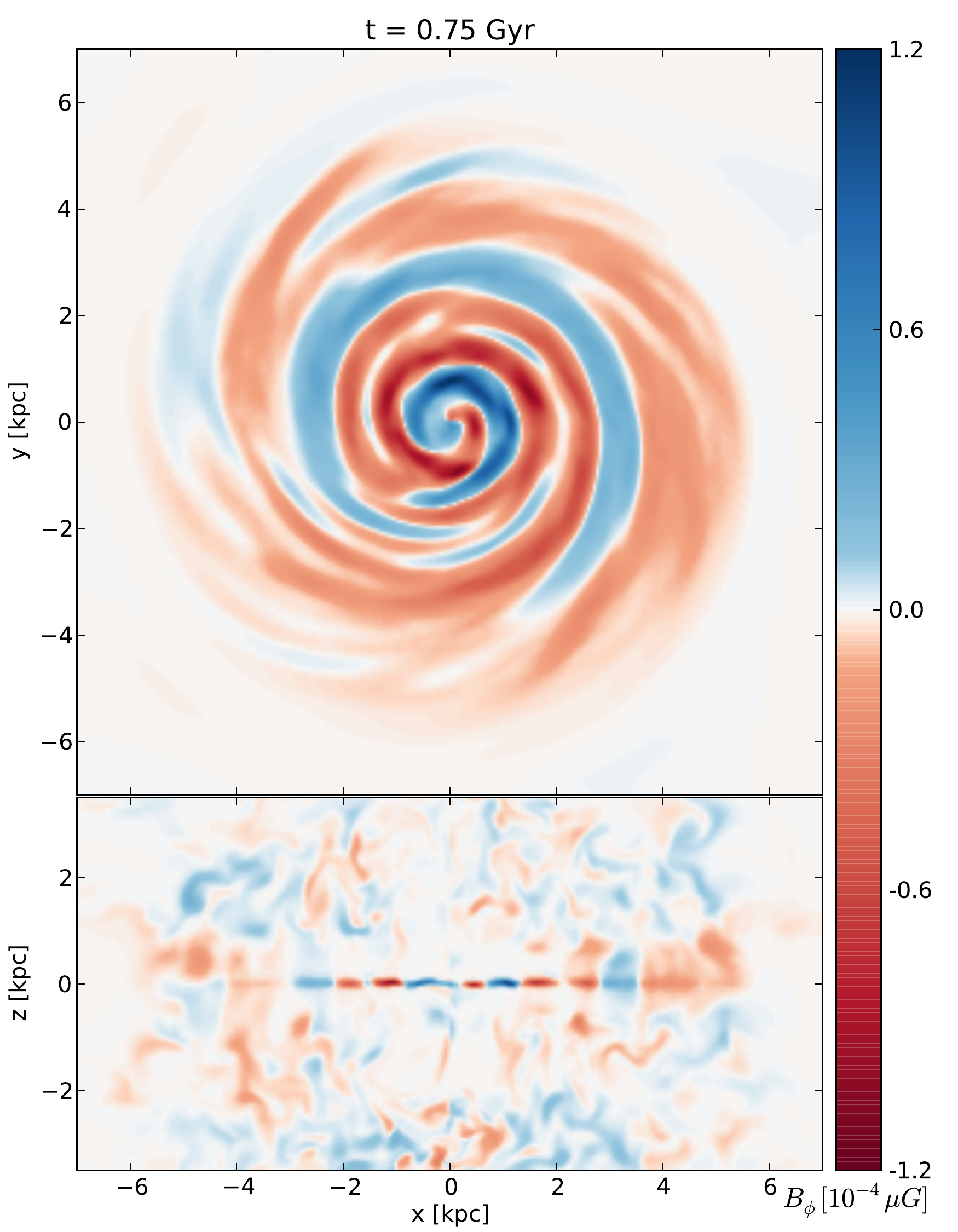}
  \includegraphics[width=0.32\textwidth]{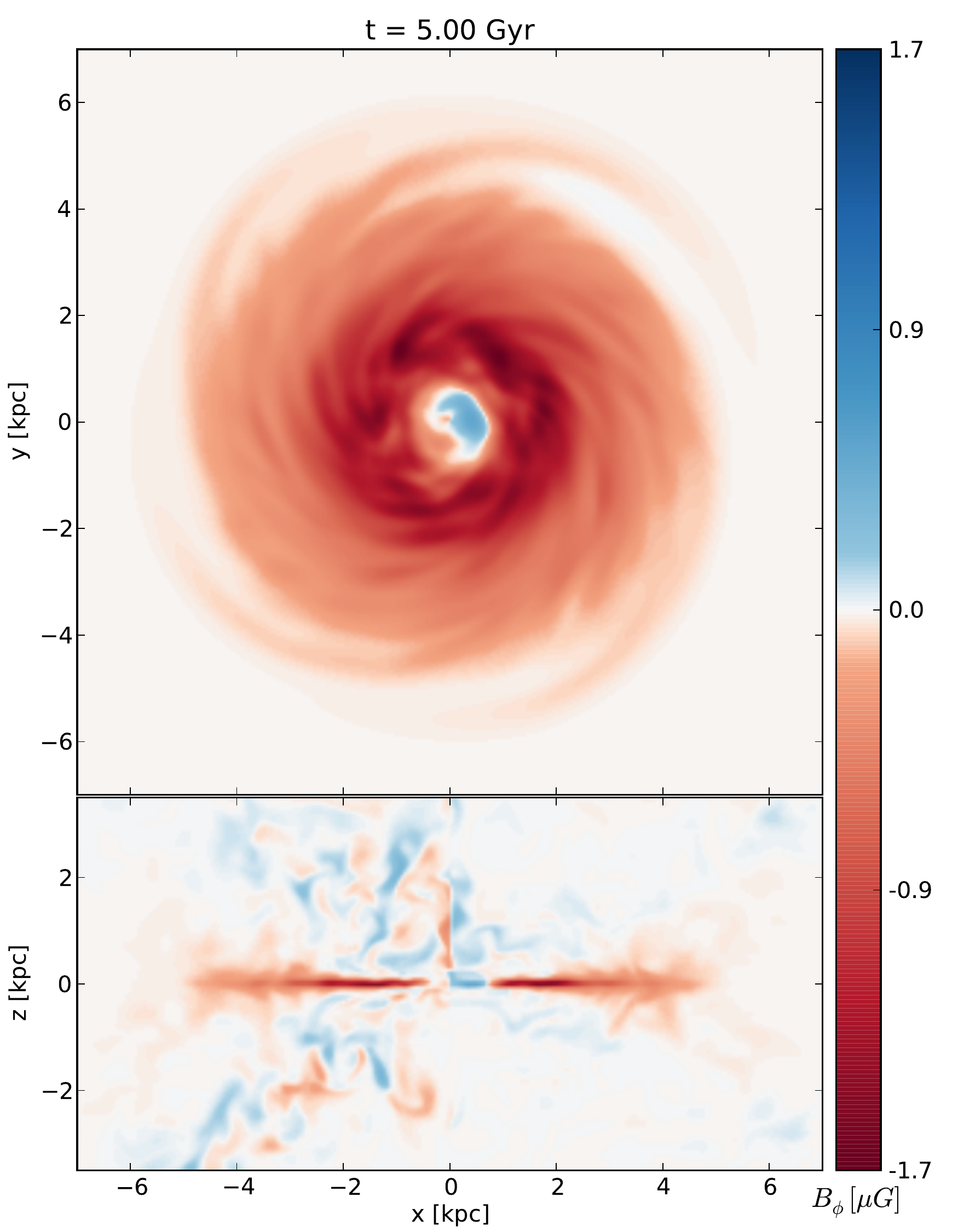}\\
  \includegraphics[width=0.32\textwidth]{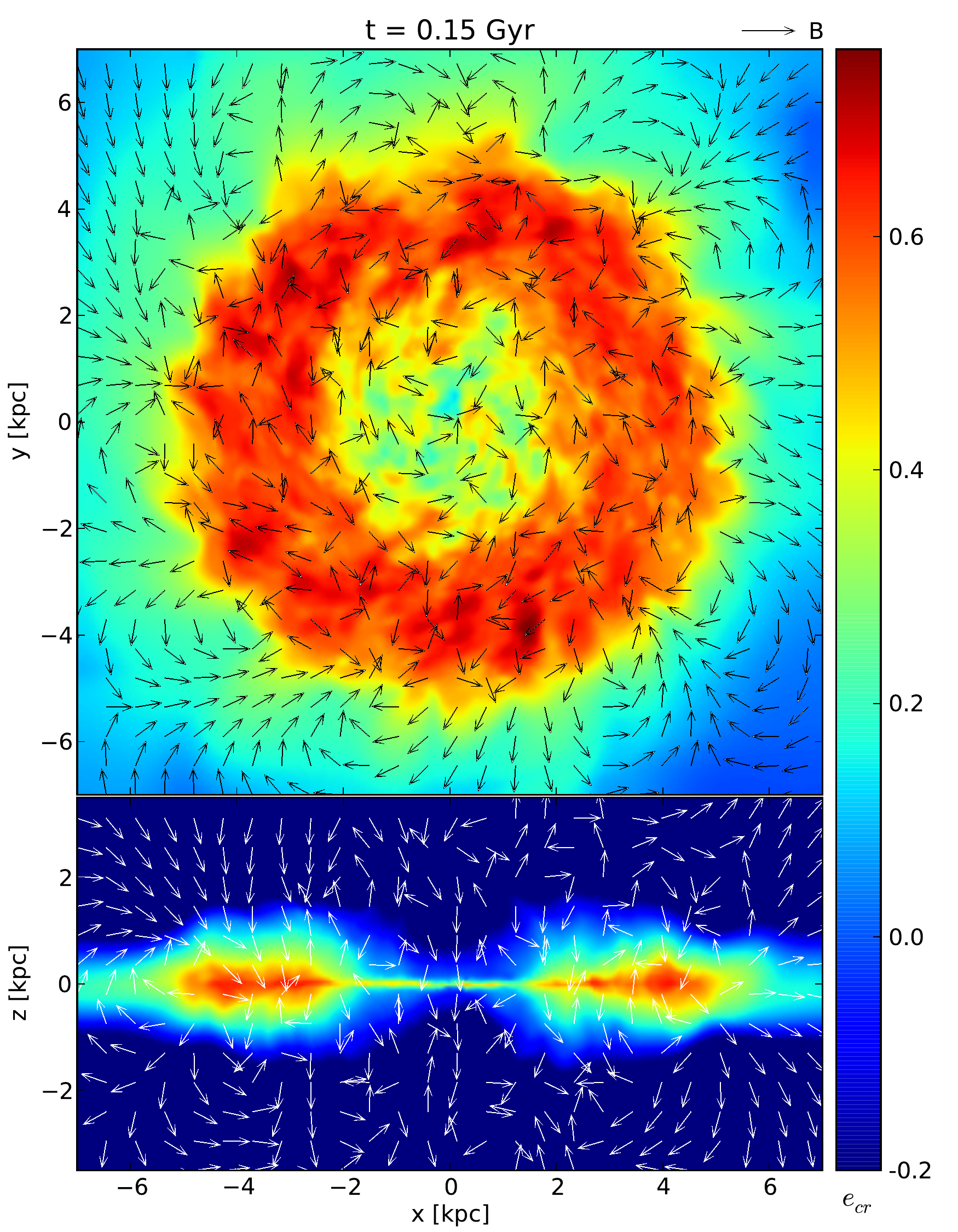}
  \includegraphics[width=0.32\textwidth]{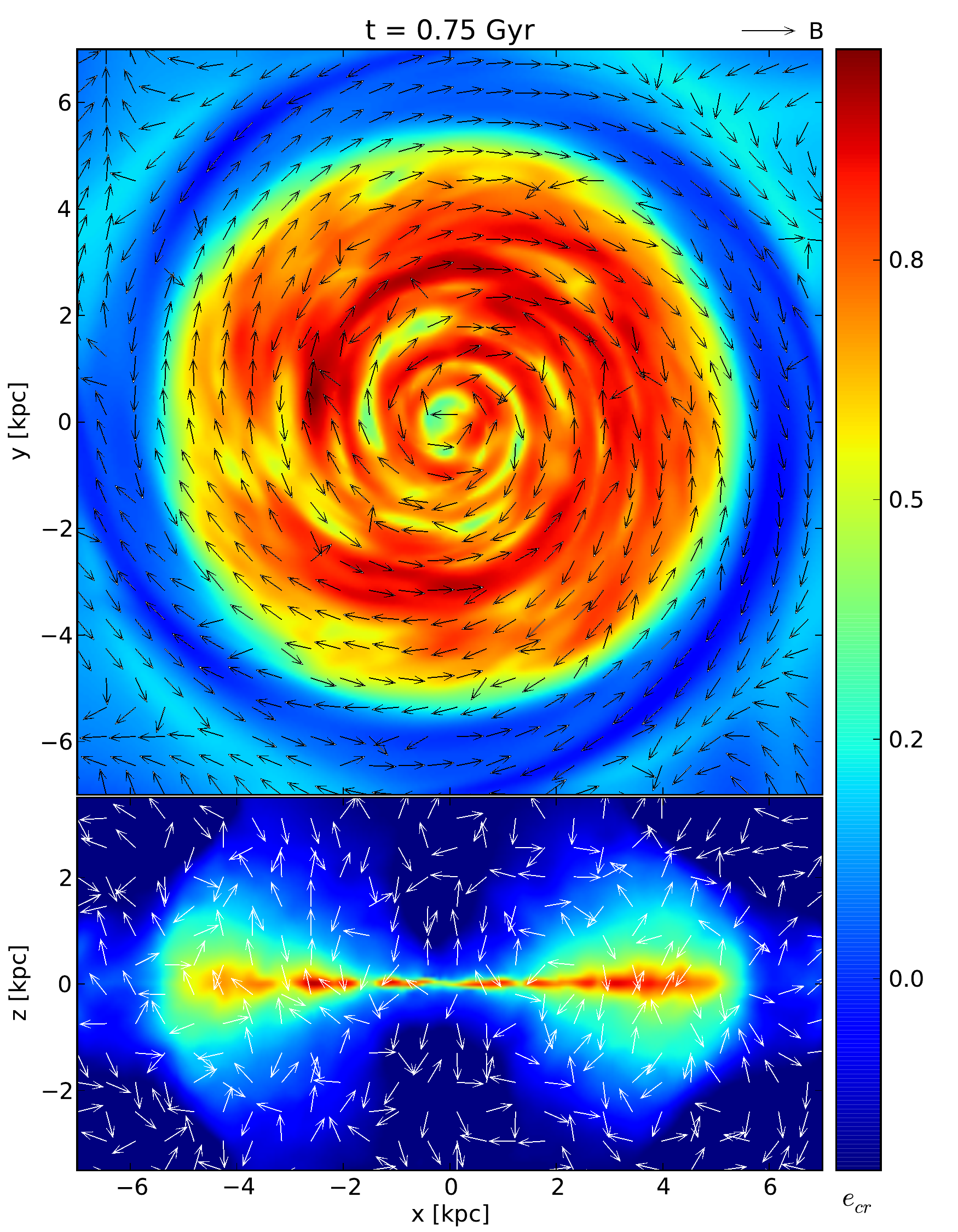}
  \includegraphics[width=0.32\textwidth]{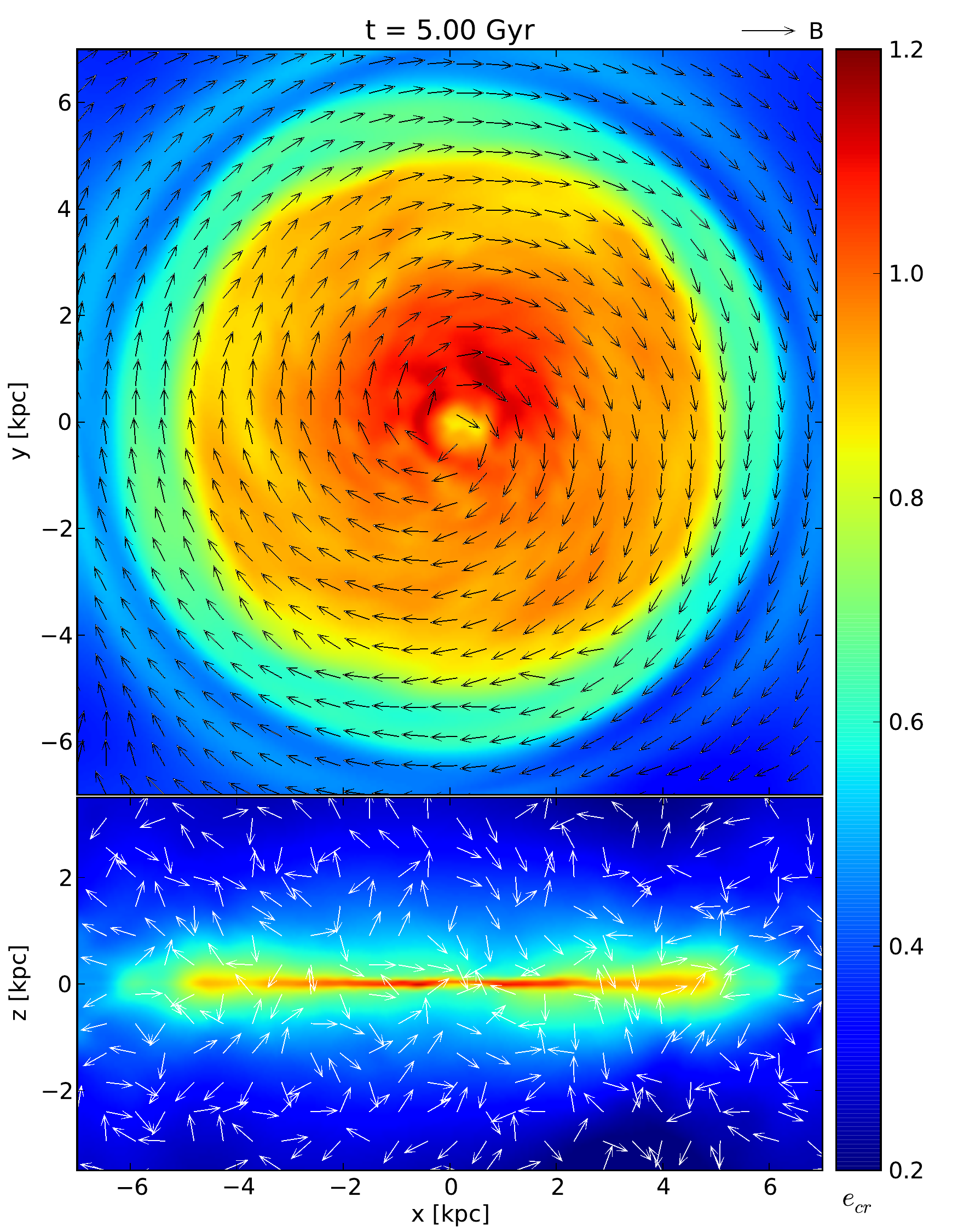}
  \caption{Evolution of the magnetic field. The subsequent upper panels show the evolution of the
      $B_\phi$ component (expressed in \muG) in edge-on and face-on view of the model v70 for
      selected time-steps: 0.15, 0.75, and 5.00~Gyr.  The lower panels show the energy density of
      the cosmic rays and the normalised vectors of the magnetic field at the corresponding
      time-steps.}
  \label{fig:maps}
\end{figure*}

In Fig.~\ref{fig:maps} the slices through the domain of model v70 are shown for selected time-steps.
The evolution of $B_\phi$ shows the initially randomly oriented magnetic field, injected through the
magnetised supernova explosions, is efficiently regularised. In the $xy$ plane at $t=0.15$~Gyr
randomly positioned spots appear of a positive (blue) and negative (red) azimuthal magnetic field.
In the $xz$ cut a similar pattern is revealed but only very close to the disk midplane, that is
there are no such structures in the halo. The differential rotation shears the radially aligned
structures and forms a spiral pattern of an oppositely directed azimuthal magnetic field, which are
clearly visible at $t = 0.75$~Gyr. In the disk at the end of the simulation there is only negatively
oriented magnetic azimuthal flux. The magnitude of $B_\phi$ is the strongest in the central part of
the disk, and it slowly decreases with the radius. The negatively oriented flux is mostly located in
the midplane of the galaxy, but it extends also to some vertical height in the form of a disk
corona.

The initial state of the cosmic-ray energy follows the density distribution.  As the system evolves
and the supernovae explode, additional energy fluctuations appear (Figure~\ref{fig:maps} bottom
panels) . Weak and wide streams of cosmic-ray energy form in the vertical direction; they start from
the disk midplane and span up to the top and bottom domain boundaries (see Figure~\ref{fig:maps} at
$t=0.75$~Gyr). These are the channels that transfer the cosmic-ray energy out of the domain.

\begin{figure}
  \includegraphics[width=0.49\textwidth]{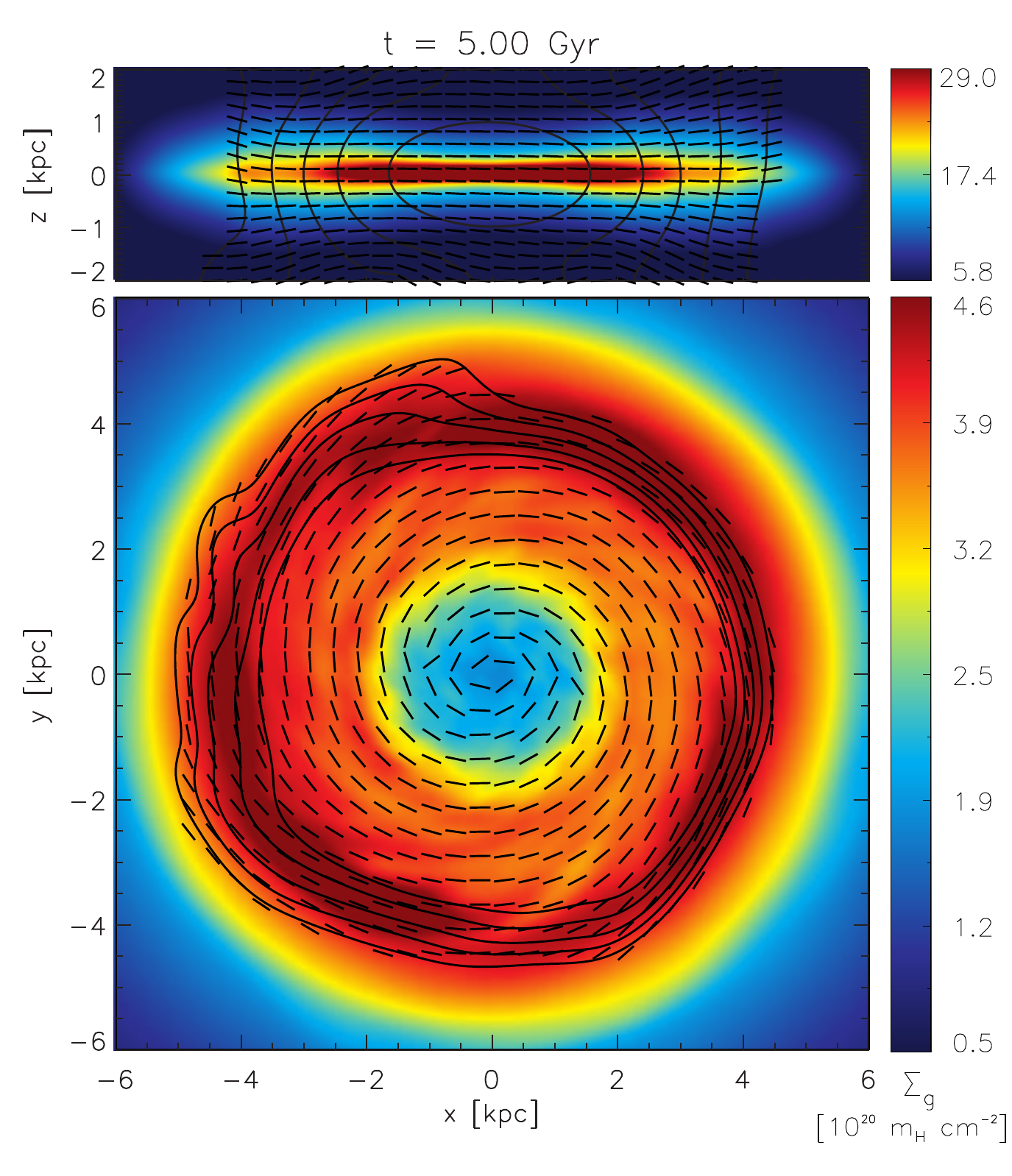}
  \caption{Polarization map at $\lambda6.2$~cm for the model v70 at $t=5$~Gyr. The top panel shows
      the edge-on and the bottom the face-on view of the galaxy. The contours of the polarisation
      intensity and the dashes of the polarisation angles are superimposed onto the column gas density
      $(\Sigma_g)$ plot, where $m_H$ is the hydrogen mass. The map have been smoothed down to the resolution 40''.
  }
  \label{fig:polar}
\end{figure}

The result of each simulation are the 3D cubes of magnetic field and the CR energy density. Using
the CR component as the proxy for the distribution of the relativistic electrons, one can create a
synthetic maps of total power and polarised synchrotron radiation \citep[for details see][Sec.
3]{otmianowska09}. Fig.~\ref{fig:polar} shows the polarisation map for model v70 at $t=5$~Gyr. The
map shows the synthetic distribution of the polarised intensity at $\lambda6.2$~cm and the
polarisation angles, both superimposed onto the column gas density. The polarised intensity in the
disk region has a very strong gradient at the edge of the disk.  The magnetic-field structure is
dominated by the azimuthal component and the vertical slices through the computational domain show
that the vertical magnetic-field component is relatively weak.  The synthetic polarisation map of
the simulated galaxy shows that the magnetic field has a very strong toroidal component and is
almost perfectly parallel to the disk plane. Similar results have been found in observation of
\object{NGC 1569} by \cite{kepley10}. This clearly hints at a long-term enhanced SFR in \object{NGC
1569} as the origin of the magnetic field, with the current burst only modulating the field
topology. The long enhanced SFR in \object{NGC 1569} is consistent with the results from HST-based
colour magnitude analyses \citep{vallenari96}.

% section results (end)

\section{Conclusions} 
\label{sec:discussion_and_conclusions}

We have shown for the first time a global model of the cosmic-ray-driven dynamo for a dwarf galaxy.
The model consists of (1) exploding supernovae that supply the CR energy input and are distributed
depending on the local gas density, (2) seeding the dynamo by randomly oriented dipoles injected
with the first bursts of supernovae, (3) the Burkert DM profile, and (4) the ISM resistivity.  We
studied the evolution of the magnetic field  for two galaxies characterised by different rotation
speeds: 40~km~s$^{-1}$ and 70~km~s$^{-1}$. We found that

\begin{itemize}

  \item the cosmic-ray-driven dynamo operating in a dwarf galaxy can amplify regular  magnetic
    fields exponentially in time, up to the saturation level that corresponds to the equipartition
    magnetic-field strength in real galaxies,

  \item the $e$-folding time-scales show that fast-rotating objects generate magnetic fields faster than
    the slow ones, which is consistent with the $\alpha\omega$-dynamo paradigm
    \citep{brandenburg05},
    
  \item the  calculated time-scales for the magnetic-field energy evolution are compatible with those
    reported by \cite{hanasz09global} and \cite{kulpa11} for more massive galaxies that show an
    $e$-folding time of the magnetic-field growth rate of about 300~Myr. In our results,
    the time-scales are about 450~Myr for v70 model and 1\,000~Myr for v40, implying that the
    magnetic field in dwarfs is more the result of a long-term slightly enhanced star formation
    than due to one recent strong burst,

  \item the 8~\muG magnetic fields generated in the model v70 are in the range of observed values
    presented in \cite{chyzy11}.  The magnetic field generated in model v40 reaches the saturation
    phase after about 10~Gyr and the final values are also similar to those of real galaxies. One
    should keep in mind that the SFR history or interaction with the environment in real galaxies
    differs from the modelled ones, therefore the magnetic-field values do not match exactly.

\end{itemize}

The results of our modelling indicate that the cosmic-ray-driven dynamo can explain the observed magnetic
fields in dwarf galaxies. In future work we plan to determine the influence of other
parameters and perform more simulations to find and reproduce the observed relations.

\begin{acknowledgements}
  This work was supported by the by the Polish National Science Centre through the grants N N203
  583440, N N203 511038, and NCN UMO-2011/03/B/ST9/01859. Calculations were made possible thanks to
  the PL-Grid Infrastructure, website: www.plgrid.pl.  This research was supported by the
  partnership program between the Jagellionian University Krak\'ow and the Ruhr-University Bochum.
  DJB is supported by the DFG special research unit FOR 1254 "Magnetisation of Interstellar and
  Intergalactic Media: The Prospects of Low-Frequency Radio Observations". We thank J. Gallagher for
  discussions.
\end{acknowledgements}

% section discussion_and_conclusions (end)

\bibliographystyle{aa}
\bibliography{refs}
  
\end{document}